# Halide mixing inhibits exciton transport in two-dimensional perovskites despite phase purity

Michael Seitz[1,2,3,4], Marc Meléndez[5], Peyton York[6], Daniel A. Kurtz[3], Alvaro J. Magdaleno[1,2], Nerea Alcázar[1,5], Mahesh K. Gangishetty[3,6], Rafael Delgado-Buscalioni[1,5], Daniel N. Congreve[3,4], and Ferry Prins[1,2]*

1. Condensed Matter Physics Center (IFIMAC), Autonomous University of Madrid, 28049 Madrid, Spain

2. Department of Condensed Matter Physics, Autonomous University of Madrid, 28049 Madrid, Spain

3. Rowland Institute at Harvard University, Cambridge, Massachusetts 02142, United States

4. Department of Electrical Engineering, Stanford University, Stanford, CA 94305, United States

5. Department of Theoretical Condensed Matter Physics, Autonomous University of Madrid, 28049 Madrid, Spain

6. Department of Chemistry, Mississippi State University, MS 39762, United States

To whom correspondence should be addressed: *ferry.prins@uam.es*

# Abstract

Metal-halide perovskites are a versatile material platform for light-harvesting and light-emitting applications as their variable chemical composition allows the optoelectronic properties to be tailored to specific applications. Halide mixing is one of the most powerful techniques to tune the optical bandgap of metal-halide perovskites across wide spectral ranges. However, halide mixing has commonly been observed to result in phase segregation, which reduces excited-state transport and limits device performance. While the current emphasis lies on the development of strategies to prevent phase segregation, it remains unclear how halide mixing may affect excited-state transport even if phase purity is maintained. In this work, we study excitonic excited-state transport in phase pure mixed-halide 2D perovskites of $(PEA)_2Pb(I_{1-x}Br_x)_4$. Using transient photoluminescence microscopy, we show that, despite phase purity, halide mixing inhibits exciton transport in these materials. We find a significant reduction even for relatively low alloying concentrations, with bromide-rich perovskites being particularly sensitive to the introduction of iodide ions. Performing Brownian dynamics simulations, we are able to reproduce our experimental results and attribute the decrease in diffusivity to the energetically disordered potential landscape that arises due to the intrinsic random distribution of alloying sites. Our results suggest that even in the absence of phase segregation, halide mixing may still impact carrier transport due to the local intrinsic inhomogeneities in the energy landscape.

# Introduction

Metal-halide perovskites have become an important material platform for light-harvesting[1–4] and light-emitting[5,6] applications thanks to their numerous advantageous properties such as solution processability, high ambipolar charge-carrier mobilities,[7,8] high defect tolerance,[9–11] and variable optical properties.[12,13] One key advantage of metal-halide perovskites is their widely tunable optical bandgap, which can be readily adjusted by introducing variations in the halide composition.[12–14] Halide mixing has been widely employed in perovskite tandem solar cells to tune the bandgap and maximize the absorptive efficiency.[15,16] In addition, halide mixing can be used in light emitting devices (LEDs) to fine-tune the color of emission – a strategy that has been successfully employed for both bulk (3D) and layered (2D) perovskites.[17,18]

However, halide mixing has been associated with reduced bandgap stability as a result of ion migration.[19,20] This instability is a result of the soft inorganic lattice of perovskites, which reduces activation energies for ion migration.[20] In extreme cases, mixed-halide perovskites have been observed to undergo phase segregation, leading to extended regions of high and low energy sites. For example, in MAPb($I_{1-x}Br_x$)$_3$ 3D perovskites, it has been found that halide mixing yields a miscibility gap in the range of 30 to 90% of bromide concentration, causing the formation of iodide-rich and bromide-rich regions.[21] In addition to this intrinsic miscibility gap, mixed-halide perovskites have been shown to suffer from light-induced phase segregation.[22–24] Crucially, phase segregation can have a significant impact on carrier transport as carriers can get trapped in low energy regions.[19,24–27] A number of strategies to avoid phase segregation in 3D perovskites are being developed, including mixed A-site cations,[19,28] or doping with tin (Sn), manganese (Mn), or potassium (K).[29,30] Interestingly, phase segregation appears to be absent in 2D perovskites.[12,31–33] This may be attributed to the increased flexibility of the 2D lattice, which is more tolerant to local strain.

Importantly though, even in the absence of phase segregation, inhomogeneous distributions of different alloy ions in the mixed-halide crystal can lead to *local* variations in the bandgap energy.[31,32]

Such site-to-site energetic disorder is caused by local statistical fluctuations in chemical composition and is commonly encountered in mixed crystal systems such as $Al_xGa_{1-x}As$, $ZnSe_xTe_{1-x}$, or $PbI_{2(1-x)}Br_{2x}$.[31,32,34–36] Indeed, Lanty et al. found that the optical properties of 2D mixed-halide perovskites $(PEA)_2Pb(I_{1-x}Br_x)_4$ are consistent with this mixed crystal picture, reproducing the change in position and width of the optical absorption spectrum for the different halide ratios by only accounting for the local statistical fluctuations.[32] Therefore, despite having a phase-pure crystal lattice, the excited-state transport properties of 2D mixed-halide perovskites may still be affected by inhomogeneities in the energy landscape. To date, however, transport studies of mixed-halide perovskites have focused mainly on the effects of phase segregation.[19,25,27] It therefore remains unclear to what extent the intrinsic local statistical fluctuations in chemical composition affect the excited state transport.

In this study, we investigate the impact of halide mixing on excited-state transport in phase-pure single-crystalline 2D metal-halide perovskites of phenethylammonium lead (iodide/bromide) $(PEA)_2Pb(I_{1-x}Br_x)_4$ (x = 0 – 100%). We show that despite the absence of phase segregation, halide mixing significantly inhibits the transport of the excitonic excited state. Specifically, we observe that for bromide concentrations of 25 – 95% the exciton transport drops by a factor of more than ten as compared to the pure phases. Using transient spectroscopy, we show that this regime of strongly decreased diffusivity coincides with a significant energetic disorder that is present in the material. Performing Brownian dynamics simulations and accounting for the statistical fluctuations in chemical composition, we are able to reproduce the spatial and spectral exciton dynamics observed in our experiments. Due to the excellent agreement between simulations and experiments, we conclude that the decrease of transport properties in mixed-halide 2D perovskites is a result of energetic disorder caused by the alloying sites. Importantly, our results show that even if phase segregation is eliminated, excited state transport in metal-halide perovskites may still be significantly affected by halide mixing.

# Results

**Transient photoluminescence microscopy (TPLM).** Single crystals of phenethylammonium lead (iodide/bromide) $(PEA)_2Pb(I_{1-x}Br_x)_4$ (x = 0 – 100%) 2D perovskites were synthesized from saturated precursor solutions (see Methods).[37,38] By varying the bromide fraction x we obtain the typical optical bandgap tuning, as evidenced by the resulting photoluminescence (PL) spectra ranging from the deep blue into the green (see Supplementary Fig. 1 and 2).[12,32] Through mechanical exfoliation, we isolate single-crystalline flakes (10 – 100 μm in lateral size), which are transferred onto a microscopy slide for investigation with an oil immersion objective. Using thick flakes provides a form of self-passivation as the surface of interest is protected by the thick crystal, providing a good diffusion barrier to oxygen and moisture for the duration of our measurements.[39]

In 2D metal-halide perovskites, the optical properties are dominated by excitonic excited states due to strong quantum and dielectric confinement effects.[40] To follow the spatiotemporal evolution of photogenerated excitons, we use transient photoluminescence microscopy (TPLM), which allows the extraction of the excitonic transport properties along the 2D inorganic plane as described previously.[39,41–43] In short, we create a narrow exciton population with a pulsed and near-diffraction limited laser diode ($\lambda_{ex}$ = 405 nm) and an oil immersion objective (N.A. = 1.3). As excitons start diffusing, the exciton population broadens with time. Using a scanning avalanche photodiode, we follow the broadening of the exciton population by tracking its PL emission *I(x,t)* (see Fig. 1a) which is proportional to the exciton density at low laser fluences. For isotropic materials such as 2D perovskites, it is sufficient to scan a 1D slice of the exciton population.[42,44] The result of such a scan is presented in Fig. 1b for $(PEA)_2PbI_4$ (x = 0%), where the broadening of the emission spot is highlighted by normalizing the emission cross-section *I(x,t)* at each point in time. At t = 0, we observe a near-diffraction limited emission spot, while at later times the emission is broadened due to the outward diffusion of excitons. The rate at which the emission broadens can be quantified using the mean-square-displacement (MSD) of the emission cross-section and can be used to determine the diffusivity $D$, which describes the speed at which excitons travel through the single-crystal (see Supplementary Note 1).[39,41,42]

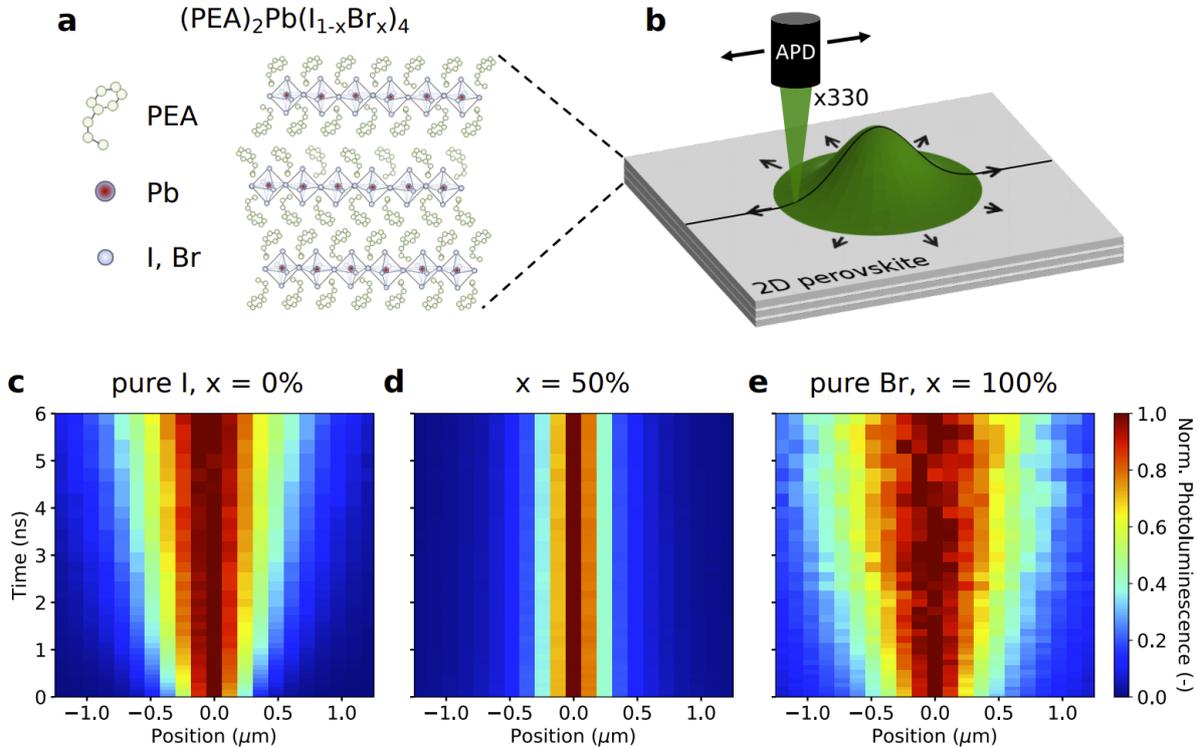

**Fig. 1 Diffusion imaging in 2D mixed-halide perovskites. a** 2D perovskite crystal structure of $(PEA)_2Pb(I_{1-x}Br_x)_4$. **b** Transient photoluminescence microscopy setup displaying the emission spot of a narrow exciton population which broadens over time as excitons diffuse outwards. The broadening is captured by magnifying the image of the emission 330 times and projecting it onto a scanning avalanche photodiode (APD) to track the evolution of excitons in space and time. **c-e** Evolution of the emission cross-section *I(x,t)* for single crystals with different halide ratios: x = 0, 50, or 100%. *I(x,t)* was normalized at each point in time to highlight the broadening of the exciton distribution.

For pure iodide $(PEA)_2PbI_4$ (x = 0%) and pure bromide $(PEA)_2PbBr_4$ (x = 100%) we find a high diffusivity with the PL cross-section quickly broadening as time goes by, indicating fast exciton diffusion (see Fig. 1c,e). Specifically, we find a comparable diffusivity for pure iodide (x = 0%, D = 0.204 cm$^2$s$^{-1}$) and pure bromide (x = 100%, D = 0.222 cm$^2$ s$^{-1}$). The diffusivity for pure iodide is consistent with our earlier studies where we synthesized single crystals with the same synthetic procedure.[39,45,46] However, when mixing equal amounts of iodide and bromide (x = 50%), we observe no broadening of the exciton distribution at all, as shown in Fig. 1d, meaning that exciton diffusion is below our detection limit of around 20 to 40 nm.

To investigate the impact of halide mixing in greater detail, we synthesize perovskite single crystals with a wider range of different bromide contents (x = 0, 5, 10, 25, 50, 75, 90, 95, and 100%). The resulting diffusion maps and MSD plots are shown in Supplementary Fig. 3-6 and the extracted diffusivities $D(x)$ and resulting diffusion length $L_D(x)$ are shown in Fig. 2. We observe that the diffusivity $D(x)$ is highly affected by halide mixing even at low levels such as 5 and 95%. In addition, mixing has a more severe impact on the bromide-rich side, where the observed diffusivity rapidly drops below our detection limit already for x = 90%, while on the iodide-rich side we still observe a measurable diffusivity for 25%. We would like to note that for the laser fluences used in this study we also exclude light-induced phase segregation as the origin of energetic disorder and reduced diffusivity, as power-dependent TPLM measurements yield results that are independent of the laser fluence (see Supplementary Note 2).

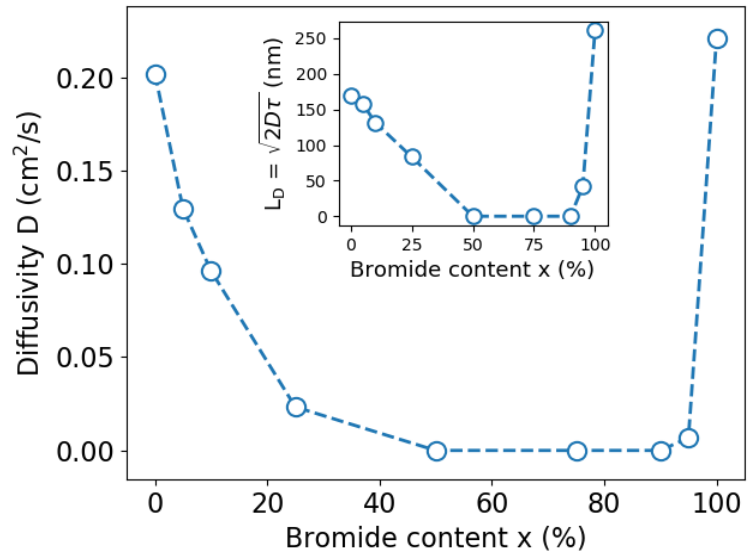

**Fig. 2 Diffusivity $D(x)$ vs. Bromide content x.** Diffusivity $D(x)$ of various 2D metal-halide perovskites (PEA)$_2$Pb(I$_{1-x}$Br$_x$)$_4$ with x = 0, 5, 10, 25, 50, 75, 90, 95, and 100%. The inset shows the diffusion length $L_D = \sqrt{2D(x)\tau(x)}$, where $\tau(x)$ is the 1/e photoluminescence lifetime (see Supplementary Fig. 7).

**Transient photoluminescence spectroscopy.** As reported in literature, significant broadening of the optical spectra is present for increased mixing in $(PEA)_2Pb(I_{1-x}Br_x)_4$ (see also Supplementary Fig. 1).[30–32,47] Lanty et al. showed that the spectral broadening in $(PEA)_2Pb(I_{1-x}Br_x)_4$ can be attributed to the presence of energetic disorder due to the local statistical fluctuation in chemical composition resulting in a higher bandgap energy in bromide-rich locations and a lower energy in iodide-rich regions, consistent with other mixed crystal systems.[31,32,34–36] To analyze the possible correlation between a decrease in diffusivity and the presence of energetic disorder we performed transient photoluminescence spectroscopy measurements using a streak camera. Fig. 3a presents normalized streak camera images for the x = 0, 50, and 100% mixed-halide perovskites, showing how the PL spectra evolve as a function of time. The dashed lines highlight the shift of the maxima, while the solid lines correspond to the evolution of the median emission energy. In Fig. 3b we show the median emission energy for all halide mixtures. We observe the largest redshifts for the mixtures which show the most dramatic decrease in diffusivity, consistent with a gradual energetically downhill migration of excitons in a disordered energy landscape. These results suggest that energetic disorder is indeed the origin of the observed trend in diffusivity. The correlation between the spatial and energetic dynamics is further emphasized by the observation of the same asymmetry in the full-width-half-max (fwhm), the energy shift, as well as the diffusivity, all of which show a bigger impact of halide mixing on the bromide-rich side (see Fig. 3c).

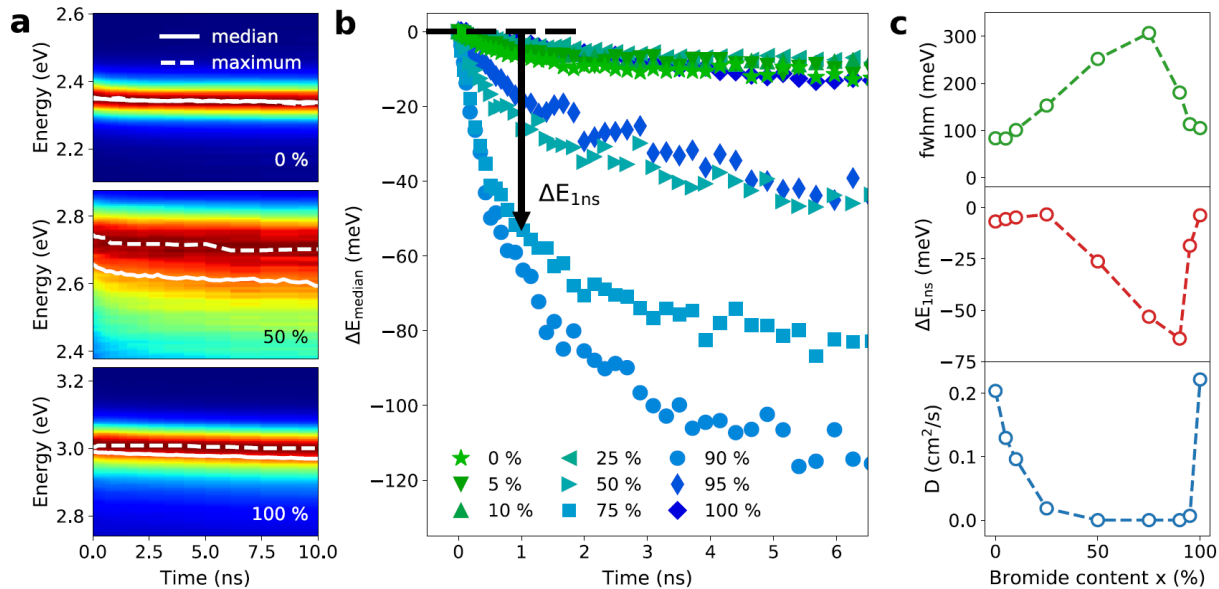

**Fig. 3 Spectrally resolved transient photoluminescence of (PEA)$_2$Pb(I$_{1-x}$Br$_x$)$_4$. a** Streak camera images for x = 0, 50, and 100% show the evolution of the spectra over 10 ns. Spectra were normalized at each point in time. The dotted line tracks the evolution of the maximum, while the solid line shows the evolution of the median emission energy. **b** Evolution of the median emission energy for various halide mixtures with x = 0, 5, 10, 25, 50, 75, 90, 95, and 100%. **c** Full-width-half-max (fwhm) of PL emission at t = 0 (top panel), median energy shift after 1 ns $\Delta E_{1ns}$ (center panel), and diffusivity D (bottom panel, same as Fig. 2) as a function for bromide content x.

**Brownian dynamics simulations.** To obtain deeper insight into the relation between exciton transport and energetic disorder due to local statistical fluctuations in the halide composition, we performed Brownian dynamics simulations. The potential landscapes $V(r)$ of the different halide mixtures are generated by randomly filling halide sites with iodide and bromide ions with a probability of (1-x) and x (see Supplementary Note 3). The probability density of a 2D exciton can be approximated with $|\Psi(R)|^2 \propto e^{-2R^2/a_B^2}$, where R is the distance from the center and $a_B$ is the exciton Bohr radius (see Supplementary Note 3).[48] Convolving the iodide and bromide sites with the exciton's probability density $|\Psi(R)|^2$, we calculate the local bromide content x'(**r**) observed by an exciton at position **r** in the crystal. Note that $x'(r)$ represents the *local* bromide content observed by the exciton at position $r$ in the crystal, while x represents the *average* bromide content of the *whole* crystal. For mixed-halide perovskites the bandgap has been shown to increase linearly with halide content.[32,49] As a result, the local energy bandgap observed by an exciton at position **r** can be calculated as $E_g(x'(r)) = (1 - x'(r)) \cdot E_g^I + x'(r) \cdot E_g^{Br}$, where $E_g^I$ and $E_g^{Br}$ are the bandgaps of the pure iodide (x = 0%) and pure bromide (x = 100%) perovskites. The resulting potential landscapes $V(r)$ ($\equiv E_g(x'(r)) - \min[E_g(x'(r))]$) for different bromide contents x are shown in Supplementary Fig. 12 and Fig. 4a, and are used for the Brownian dynamics simulations (see Supplementary Note 4). Note that we use $V(r)$ for a better visualization and comparison of the different compositions x, instead of $E_g(r)$. For the Brownian dynamics simulations we used Bohr radii that were previously reported for pure iodide ($a_B^I \equiv a_B(x = 0\%) = 1.15\ nm$) and pure bromide ($a_B^{Br} \equiv a_B(x = 100\%) = 0.7\ nm$) and linear interpolation for mixed components: $a_B(x) = (1 - x) \cdot a_B^I + x \cdot a_B^{Br}$.[14] In addition, we approximate the diffusion coefficient $D_0$ of mixed-halide perovskites, which describes the intrinsic diffusivity in the absence of any energetic disorder through linear interpolation between the two pure components: $D_0(x) = (1 - x) \cdot D_0^I + x \cdot D_0^{Br}$, with $D_0^I = D(x = 0\%) = 0.204\ cm^2 s^{-1}$ and $D_0^{Br} = D(x = 100\%) = 0.222\ cm^2 s^{-1}$ (cf. Fig. 2).

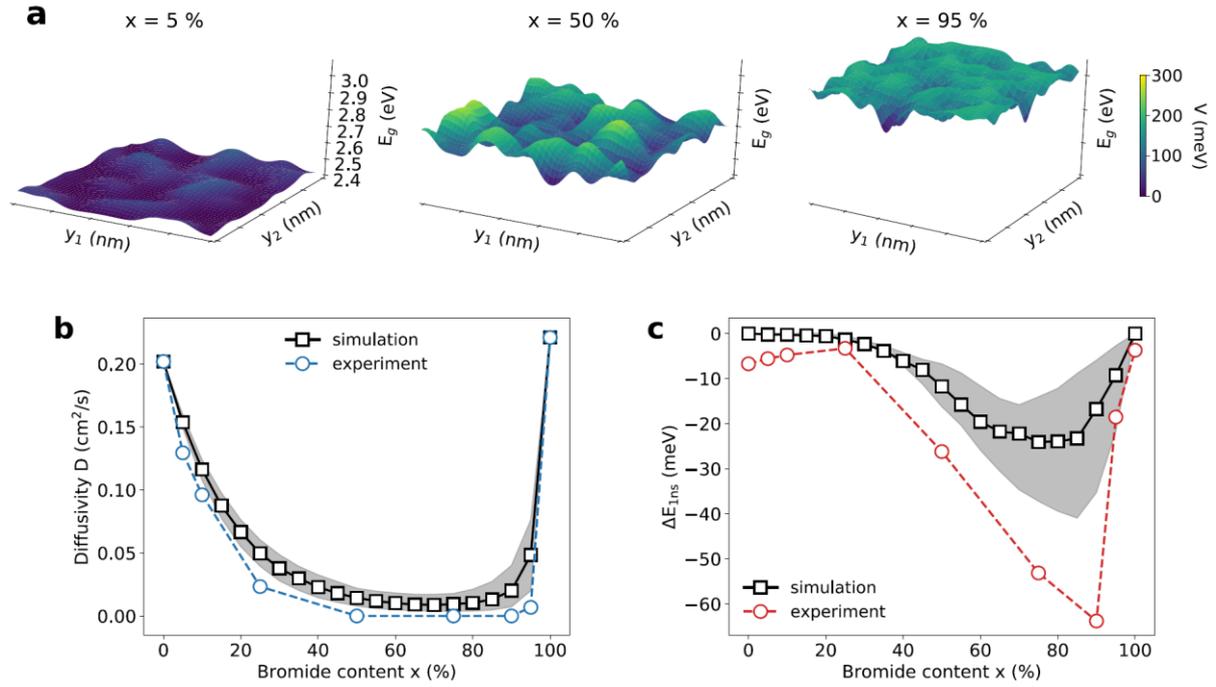

**Fig. 4 Experiments vs. Brownian dynamics simulations. a** Potential landscape $V(y_1, y_2)$ for x = 5, 50, and 95% on a 10 x 10 nm area. **b,c** Comparison of experiments (open circles) and Brownian dynamics simulations (open squares) with an exciton Bohr radius of $a_B(x) = (1-x) \cdot a_B^I + x \cdot a_B^{Br}$ ($a_B^I = 1.15\ nm$, $a_B^{Br} = 0.7\ nm$). Shaded area shows simulation results with $a_B^I = 1.15 \pm 0.1\ nm$ and $a_B^{Br} = 0.7 \pm 0.1\ nm$. **b** Diffusivity $D(x)$ as a function of bromide content x. **c** Change in median energy after 1 ns of photoexcitation $\Delta E_{1ns}$. We used a moving average ($f(x_i) = \frac{f(x_{i-1}) + f(x_i) + f(x_{i+1})}{3}$) for the simulated data points in **c**.

The results of the Brownian dynamics simulations are shown in Fig. 4b and c. We find excellent agreement with our experimental results, even in the absence of any fit parameters. The simulations also reproduce the asymmetry of lower diffusivities toward the bromide side. This asymmetry can be understood by thinking about the dilute case. Exchanging some iodide with bromide ions in a (PEA)$_2$PbI$_4$ crystal (x = 0%) will result in a flat energy landscape with an occasional elevation in the potential V(**r**) (see Fig. 4a). Excitons encountering such an elevation will not be slowed down significantly as they simply scatter away from the obstacle or move around it. On the other hand, introducing iodide into a (PEA)$_2$PbBr$_4$ (x = 100%) lattice will result in valleys in the energetic landscape, where excitons can get stuck for a certain amount of time (see Fig. 4a). Consequently, the impact of halide mixing on the transport properties is stronger on the bromide-rich side. Additionally, the asymmetry is enhanced by the smaller exciton Bohr radius $a_B^{Br} < a_B^I$, which makes excitons on the bromide-rich side more susceptible to the local fluctuations in composition x'. However, as shown in Supplementary Fig. 13, the asymmetry is clearly present even when assuming a constant exciton Bohr radius $a_B^I = a_B^{Br} = a_B(x) = 0.8\ nm$.

Brownian dynamics simulations also allow the extraction of the transient red shifts (see Supplementary Fig. 16). As shown in Fig. 4c, the simulations also agree well with our transient spectroscopy measurements from Fig. 3, reproducing the observed asymmetry and showing a good quantitative agreement for low alloying concentrations. It is worth highlighting that experimentally we observe a larger energetic red-shift in the range of x = 50 - 90% than suggested by the Brownian dynamics simulations. This could be due to uncertainties in our model, such as the exciton Bohr radius, which has a particularly strong impact on the energetic red-shift in this range, or the approximation of the exciton probability density with a Gaussian function (see Supplementary Note 3). Moreover, additional energetic disorder might be introduced by other alloying effects (e.g. strain), trap-states, or self-trapped excitons, which are not included in our model.

In conclusion, we have shown that halide mixing diminishes exciton transport even in phase pure 2D perovskites, despite the absence of phase segregation. We find that the reduced diffusivity in 2D mixed-halide perovskites is a result of the intrinsic energetic disorder, caused by the *local* statistical fluctuations in halide composition, which restricts the movement of excitons. Our measurements show a significant reduction of exciton transport even at low levels of halide mixing. This effect is particularly strong on the bromide-rich side, where the diffusivity drops by more than an order of magnitude already for 5% or iodide alloying. Our experimental observations are supported by Brownian dynamics simulations, which successfully reproduce the spatial and energetic dynamics by only accounting for the presence of local statistical fluctuations in the halide composition. Our results highlight the importance of halide mixing on the spatial excited state dynamics in 2D metal-halide perovskites and should therefore be carefully considered in the design of optimized optoelectronic devices such as solar cells. For LEDs, however, exciton funneling and concentration at lower energy sites present an opportunity for improved LED performance as the increased local exciton concentrations allow radiative recombination to outperform trap-state mediated nonradiative recombination.[50] On a final note, it is worth mentioning that most efforts to mitigate the negative impact of halide mixing in 3D perovskites have focused on the elimination of phase segregation. [12,31–33] Crucially though, our results suggest that even if phase segregation is eliminated, halide mixing may still impact carrier transport due to the local intrinsic inhomogeneities in the energy landscape.

## Methods

**Material Synthesis.** 2D perovskites of $(PEA)_2Pb(I_{1-x}Br_x)_4$ were grown following a simple supersaturation procedure under ambient conditions.[37,38] First, two 0.2 M halide-pure stock solutions of $(PEA)_2PbX_4$ (X = I or Br) were prepared by mixing PEAX (Sigma Aldrich; 805904-25G, 900829-10G) and $PbX_2$ (Sigma Aldrich; 900168-5G, 398853-5g) in a stoichiometric ratio (2:1) and dissolving the precursors in a 50/50 mixture of γ-butyrolactone and dimethyl sulfoxide (DMSO). The solutions were heated to 70°C and stirred to accelerate the dissolution of the precursors. DMSO was needed to dissolve the $PbBr_2$. Second, the I and Br stock solutions were mixed in a (1-x)/x ratio to obtain the final solutions of $(PEA)_2Pb(I_{1-x}Br_x)_4$. The $(PEA)_2Pb(I_{1-x}Br_x)_4$ solutions were dropcast onto a glass slide and after 1-3 days, mm-sized single-crystals formed, which were exfoliated using the scotch tape method and transferred to a cover slip for inspection.[39]

**Transient Photoluminescence Microscopy (TPLM).** TPLM measurements were performed following the procedure described by Akselrod et al.[41,42] In short, a near diffraction limited exciton population was created using a 405 nm pulsed laserdiode (PicoQuant LDH-D-C-405, PDL 800-D; 40 MHz, 50 nJcm$^{-2}$). The emission from the exciton population was optically magnified 330 times (Nikon CFI Plan Fluor, NA = 1.3) and imaged with a scanning avalanche photodiode (APD, Micro Photon Devices PDM, 20x20 μm detector). The laserdiode and APD were synchronized using a timing board for time correlated single photon counting (Pico-Harp 300). An x-y-piezo stage (MCL Nano-BIOS 100) was used to scan the sample during the measurement, covering an area of 5x5 μm, to reduce photodegradation of the perovskite flakes.

**Transient Photoluminescence Spectroscopy (TPLS).** TPLS measurements were performed with a Hamamatsu C10627 streak unit, which was coupled with a Hammamatsu C9300 digital camera, and a SP2150i spectrograph (Princeton Instruments). Samples were excited with a 379 nm pulsed laserdiode (Hammamatsu, 81 ps pulse width, 5 MHz, < 5 nJcm$^{-2}$).

**Brownian Dynamics Simulations.** We modeled the motion of excitons as the diffusion of independent Brownian walkers satisfying the stochastic differential equation $\Delta \boldsymbol{r} = \frac{D}{k_B T}\boldsymbol{F}\Delta t + \sqrt{2D_0}d\boldsymbol{W}$ in the standard Itô interpretation, where $F = -\nabla V$ stands for the force felt by an exciton, $D_0$ is the diffusion coefficient, $k_B T$ the thermal energy, and $d\boldsymbol{W}$ satisfies the Wiener process relation $\langle d\boldsymbol{W} d\boldsymbol{W}\rangle = \Delta t$. We integrated the equations numerically with the popular Euler-Maruyama scheme.

## Data availability

The data supporting the findings of this study are available within the article and its Supplementary Information. Additional data is available upon reasonable request to the corresponding author.

## Code availability

Correspondence and requests for codes used in the paper should be addressed to the corresponding author.

## References

1. Kojima, A., Miyasaka, T., Teshima, K., Shirai, Y. & Miyasaka, T. Organometal halide perovskites as visible-light sensitizers for photovoltaic cells. *J. Am. Chem. Soc.* **131**, 6050–6051 (2009).
2. Burschka, J. *et al.* Sequential deposition as a route to high-performance perovskite-sensitized solar cells. *Nature* **499**, 316–319 (2013).
3. Liu, M., Johnston, M. B. & Snaith, H. J. Efficient planar heterojunction perovskite solar cells by

vapour deposition. *Nature* **501**, 395–398 (2013).

4. Green, M. A., Ho-Baillie, A. & Snaith, H. J. The emergence of perovskite solar cells. *Nat. Photonics* **8**, 506–514 (2014).

5. Veldhuis, S. A. *et al.* Perovskite Materials for Light-Emitting Diodes and Lasers. *Adv. Mater.* **28**, 6804–6834 (2016).

6. Tan, Z.-K. *et al.* Bright light-emitting diodes based on organometal halide perovskite. *Nat. Nanotechnol.* **9**, 687–692 (2014).

7. Stranks, S. D. *et al.* Electron-hole diffusion lengths exceeding 1 micrometer in an organometal trihalide perovskite absorber. *Science* **342**, 341–344 (2013).

8. Shi, D. *et al.* Low trap-state density and long carrier diffusion in organolead trihalide perovskite single crystals. *Science* **347**, 519–522 (2015).

9. Yin, W. J., Shi, T. & Yan, Y. Unusual defect physics in $CH_3NH_3PbI_3$ perovskite solar cell absorber. *Appl. Phys. Lett.* **104**, (2014).

10. Brandt, R. E., Stevanović, V., Ginley, D. S. & Buonassisi, T. Identifying defect-tolerant semiconductors with high minority-carrier lifetimes: Beyond hybrid lead halide perovskites. *MRS Commun.* **5**, 265–275 (2015).

11. Steirer, K. X. *et al.* Defect Tolerance in Methylammonium Lead Triiodide Perovskite. *ACS Energy Lett.* **1**, 360–366 (2016).

12. Weidman, M. C., Seitz, M., Stranks, S. D. & Tisdale, W. A. Highly Tunable Colloidal Perovskite Nanoplatelets through Variable Cation, Metal, and Halide Composition. *ACS Nano* **10**, 7830–7839 (2016).

13. Shamsi, J., Urban, A. S., Imran, M., De Trizio, L. & Manna, L. Metal Halide Perovskite Nanocrystals: Synthesis, Post-Synthesis Modifications, and Their Optical Properties. *Chem. Rev.* **119**, 3296–3348 (2019).

14. Papavassiliou, G. C. Three- and Low-Dimensional Inorganic Semiconductors. in *Progress in Solid State Chemistry* **25**, 125–270 (Elsevier Ltd, 1997).

15. McMeekin, D. P. *et al.* A mixed-cation lead mixed-halide perovskite absorber for tandem solar cells. *Science* **351**, 151–155 (2016).

16. Jošt, M., Kegelmann, L., Korte, L. & Albrecht, S. Monolithic Perovskite Tandem Solar Cells: A Review of the Present Status and Advanced Characterization Methods Toward 30% Efficiency. *Advanced Energy Materials* **10**, 1904102 (2020).

17. Xiao, Z. *et al.* Mixed-Halide Perovskites with Stabilized Bandgaps. *Nano Lett.* **17**, 6863–6869 (2017).

18. Hassan, Y. *et al.* Ligand-engineered bandgap stability in mixed-halide perovskite LEDs. *Nature* **591**, 72–77 (2021).

19. Rehman, W. *et al.* Photovoltaic mixed-cation lead mixed-halide perovskites: Links between crystallinity, photo-stability and electronic properties. *Energy Environ. Sci.* **10**, 361–369 (2017).

20. Nah, Y. *et al.* Spectral Instability of Layered Mixed Halide Perovskites Results from Anion Phase Redistribution and Selective Hole Injection. *ACS Nano* **15**, 1486–1496 (2021).

21. Lehmann, F. *et al.* The phase diagram of a mixed halide (Br, I) hybrid perovskite obtained by synchrotron X-ray diffraction. *RSC Adv.* **9**, 11151–11159 (2019).

22. Knight, A. J. *et al.* Electronic Traps and Phase Segregation in Lead Mixed-Halide Perovskite. *ACS Energy Lett.* **4**, 75–84 (2019).

23. Brennan, M. C., Draguta, S., Kamat, P. V. & Kuno, M. Light-Induced Anion Phase Segregation in Mixed Halide Perovskites. *ACS Energy Lett.* **3**, 204–213 (2018).

24. Hoke, E. T. *et al.* Reversible photo-induced trap formation in mixed-halide hybrid perovskites for photovoltaics. *Chem. Sci.* **6**, 613–617 (2015).

25. Unger, E. L. *et al.* Roadmap and roadblocks for the band gap tunability of metal halide perovskites. *J. Mater. Chem. A* **5**, 11401–11409 (2017).

26. Mahesh, S. *et al.* Revealing the origin of voltage loss in mixed-halide perovskite solar cells. *Energy Environ. Sci.* **13**, 258–267 (2020).

27. Diez-Cabanes, V., Even, J., Beljonne, D. & Quarti, C. Electronic Structure and Optical Properties of Mixed Iodine/Bromine Lead Perovskites. To Mix or Not to Mix? *Adv. Opt. Mater.* 2001832 (2021). doi:10.1002/adom.202001832

28. Correa-Baena, J.-P. *et al.* Homogenized halides and alkali cation segregation in alloyed organic-inorganic perovskites. *Science* **363**, 627–631 (2019).

29. Hong, D. *et al.* Inhibition of Phase Segregation in Cesium Lead Mixed-Halide Perovskites by B-Site Doping. *iScience* **23**, 101415 (2020).

30. Ahmad, S., Rahil, M., Rajput, P. & Ghosh, D. Highly tunable single-phase excitons in mixed halide layered perovskites. *ACS Appl. Electron. Mater.* **2**, 3199–3210 (2020).

31. Ahmad, S., Baumberg, J. J. & Vijaya Prakash, G. Structural tunability and switchable exciton emission in inorganic-organic hybrids with mixed halides. *J. Appl. Phys.* **114**, 233511 (2013).

32. Lanty, G. *et al.* Room-Temperature optical tunability and inhomogeneous broadening in 2D-layered organic-inorganic perovskite pseudobinary alloys. *J. Phys. Chem. Lett.* **5**, 3958–3963 (2014).

33. Zhang, X. *et al.* Perovskite (PEA)2Pb(I1-xBrx)4 single crystal thin films for improving optoelectronic performances. *Opt. Mater. (Amst).* **117**, 111074 (2021).

34. Takeda, J., Tayu, T., Saito, S. & Kurita, S. Exciton-Phonon Interaction and Potential Fluctuation Effect in PbI$_{2(1-x)}$Br$_{2x}$ Mixed Crystals. *J. Phys. Soc. Japan* **60**, 3874–3881 (1991).

35. Naumov, A., Stanzl, H., Wolf, K., Lankes, S. & Gebhardt, W. Exciton recombination in Te-rich ZnSexTe1-x epilayers. *J. Appl. Phys.* **74**, 6178–6185 (1993).

36. Lee, S. M. & Bajaj, K. K. A quantum statistical theory of linewidths of radiative transitions due to compositional disordering in semiconductor alloys. *J. Appl. Phys.* **73**, 1788–1796 (1993).

37. Seitz, M., Gant, P., Castellanos-Gomez, A. & Prins, F. Long-Term Stabilization of Two-Dimensional Perovskites by Encapsulation with Hexagonal Boron Nitride. *Nanomaterials* **9**, 1120 (2019).

38. Ha, S. T., Shen, C., Zhang, J. & Xiong, Q. Laser cooling of organic-inorganic lead halide perovskites. *Nat. Photonics* **10**, 115–121 (2016).

39. Seitz, M. *et al.* Exciton diffusion in two-dimensional metal-halide perovskites. *Nat. Commun.* **11**, (2020).

40. Mauck, C. M. & Tisdale, W. A. Excitons in 2D Organic–Inorganic Halide Perovskites. *Trends Chem.* **1**, 380–393 (2019).

41. Akselrod, G. M. *et al.* Visualization of exciton transport in ordered and disordered molecular solids. *Nat. Commun.* **5**, 3646 (2014).

42. Akselrod, G. M. *et al.* Subdiffusive exciton transport in quantum dot solids. *Nano Lett.* **14**, 3556–3562 (2014).

43. Ziegler, J. D. *et al.* Fast and Anomalous Exciton Diffusion in Two-Dimensional Hybrid Perovskites. *Nano Lett.* **20**, 6674–6681 (2020).

44. Deng, S. *et al.* Long-range exciton transport and slow annihilation in two-dimensional hybrid perovskites. *Nat. Commun.* **11**, 1–8 (2020).

45. Seitz, M. *et al.* Mapping the Trap-State Landscape in 2D Metal-Halide Perovskites Using Transient Photoluminescence Microscopy. *Adv. Opt. Mater.* 2001875 (2021). doi:10.1002/adom.202001875

46. Magdaleno, A. J. *et al.* Efficient interlayer exciton transport in two-dimensional metal-halide perovskites. *Mater. Horizons* **8**, 639–644 (2021).

47. Congreve, D. N. *et al.* Tunable Light-Emitting Diodes Utilizing Quantum-Confined Layered Perovskite Emitters. *ACS Photonics* **4**, 476–481 (2017).

48. Prada, E., Alvarez, J. V., Narasimha-Acharya, K. L., Bailen, F. J. & Palacios, J. J. Effective-mass theory for the anisotropic exciton in two-dimensional crystals: Application to phosphorene. *Phys. Rev. B* **91**, 245421 (2015).

49. Weidman, M. C., Seitz, M., Stranks, S. D. & Tisdale, W. A. Highly Tunable Colloidal Perovskite Nanoplatelets through Variable Cation, Metal, and Halide Composition. *ACS Nano* **10**, 7830–7839 (2016).

50. Yuan, M. *et al.* Perovskite energy funnels for efficient light-emitting diodes. *Nat. Nanotechnol.* **11**, 872–877 (2016).

## Acknowledgment

We thank José Vicente Álvarez and Elsa Prada for helpful discussions. This work has been supported by the Spanish Ministry of Economy and Competitiveness through the "María de Maeztu" Program for Units of Excellence in R&D (MDM-2014-0377). M.S. acknowledges the financial support through a Doc.Mobility Fellowship from the Swiss National Science Foundation (SNF) with grant number 187676. In addition, M.S. acknowledges the financial support of a fellowship from "la Caixa" Foundation (ID 100010434). The fellowship code is LCF/BQ/IN17/11620040. Further, M.S. has received funding from the European Union's Horizon 2020 research and innovation program under the Marie Skłodowska-Curie grant agreement No. 713673. F.P. acknowledges support from the Spanish Ministry for Science, Innovation, and Universities through the state program (PGC2018-097236-A-I00) and through the Ramón y Cajal program (RYC-2017-23253), as well as the Comunidad de Madrid Talent Program for Experienced Researchers (2016-T1/IND-1209). M.M., N.C., and R.D.B. acknowledge support from the Spanish Ministry of Economy, Industry, and Competitiveness through Grant FIS2017-86007-C3-1-P (AEI/FEDER, EU). D.N.C acknowledges the support of the Rowland Fellowship at the Rowland Institute at Harvard University and the Department of Electrical Engineering at Stanford University. M.K.G. acknowledges the support of National Science Foundation Track 1 EPSCoR funding under the Grant No. #1757220. D.A.K. acknowledges the support of a Rowland Foundation Postdoctoral Fellowship.

## Author contributions

M.S. and F.P. designed this study. M.S. led the experimental work and processing of experimental data. M.S., P.Y., and A.J.M. prepared perovskite materials. M.S., M.K.G., P.Y., D.K., and A.J.M. performed material characterization. M.M., M.S., N.A., and R.D.-B. performed theoretical and numerical modeling of exciton transport. F.P. and D.N.C. supervised the project. M.S. and F.P. wrote the original draft of the paper. All authors contributed to reviewing the paper.

## Competing Interests

The authors declare no competing interests.

Supporting Information

# Halide alloying inhibits exciton transport in two-dimensional metal-halide perovskites

Seitz et al.

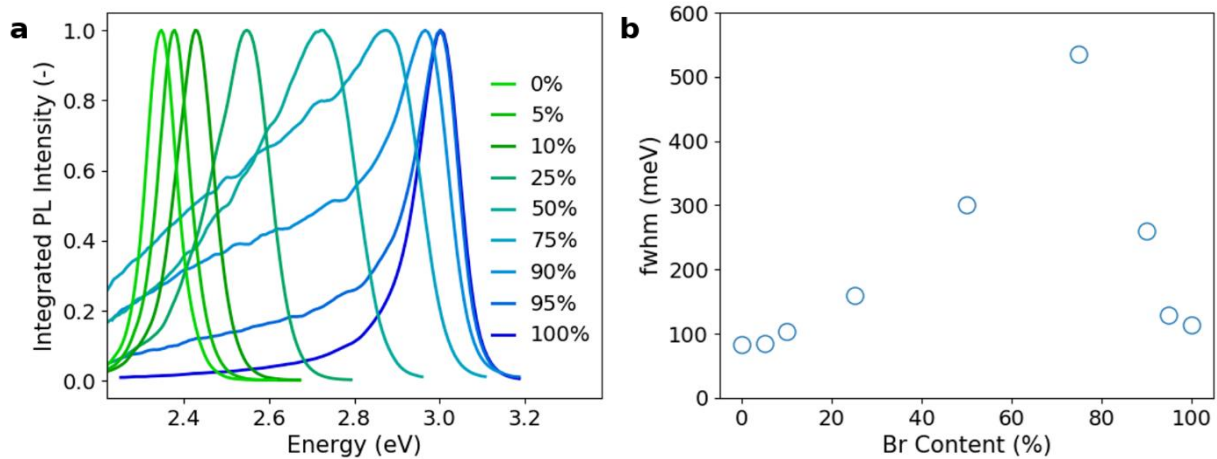

**Supplementary Fig. 1. a** Integrated photoluminescence (PL) emission from pulsed laser excitation of 2D metal halides (PEA)Pb(I$_{1-x}$Br)$_4$ with various bromide contents x = 0, 5, 10, 25, 50, 75, 90, 95, 100%. **b** Full-width-half-max (fwhm) of the PL spectra in **a**.

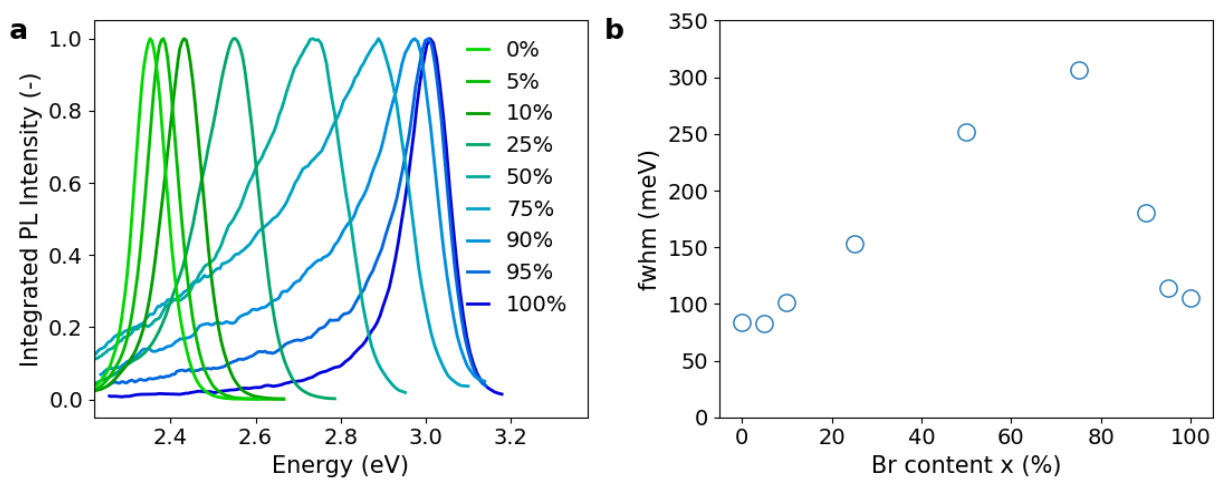

**Supplementary Fig. 2. a** Photoluminescence (PL) emission right after laser excitation (t = 0) of 2D metal halides (PEA)Pb(I$_{1-x}$Br)$_4$ with various bromide contents x = 0, 5, 10, 25, 50, 75, 90, 95, 100%. **b** Full-width-half-max (fwhm) of the PL spectra in **a**.

## Supplementary Note 1

**Transient photoluminescence microscopy (TPLM).** TPLM was performed as previously reported.[1] In short, a near-diffraction limited exciton distribution was created using a ×100 oil immersion objective (Nikon CFI Plan Fluor, NA = 1.3) and a 405 nm pulsed laser diode (PicoQuant LDH-D-C-405, PDL 800-D). The photoluminescence of the exciton population was separated from the excitation light with a 420 nm dichroic mirror and imaged onto an avalanche photodiode (APD, Micro Photon Devices PDM, 20 µm detector size) with a ×330 magnification. APD and laser were synchronized using a timing board for time-correlated single photon counting (Pico-Harp 300). During TPLM the laser was scanned over a 5 × 5 µm area of the single-crystal (Mad City Labs Nano BIOS 100 x-y-piezo stage) to minimize photodegradation. The laser repetition rate was 40 MHz and the laser fluence was 50 nJcm$^{-2}$ unless stated otherwise. The time binning of the measurement setup was set to 4 ps before software binning was applied. The resulting diffusion maps are shown in Supplementary Fig. 3. We follow the same fitting procedure as described previously by our group to extract the evolution of the mean-square-displacement (MSD) of the exciton population.[1] As shown in Supplementary Fig. 4-6, the MSD first grows linearly and then transitions to a sublinear growth (t > 1 ns) due to excitons getting stuck at trapping sites.[1] We extract the diffusivity $D(x)$ of the different alloys using the one-dimensional diffusion equation and fitting the data at early times with $MSD(t) = 2D(x)t$ (see fit results in Fig. 2 of the main text).[1]

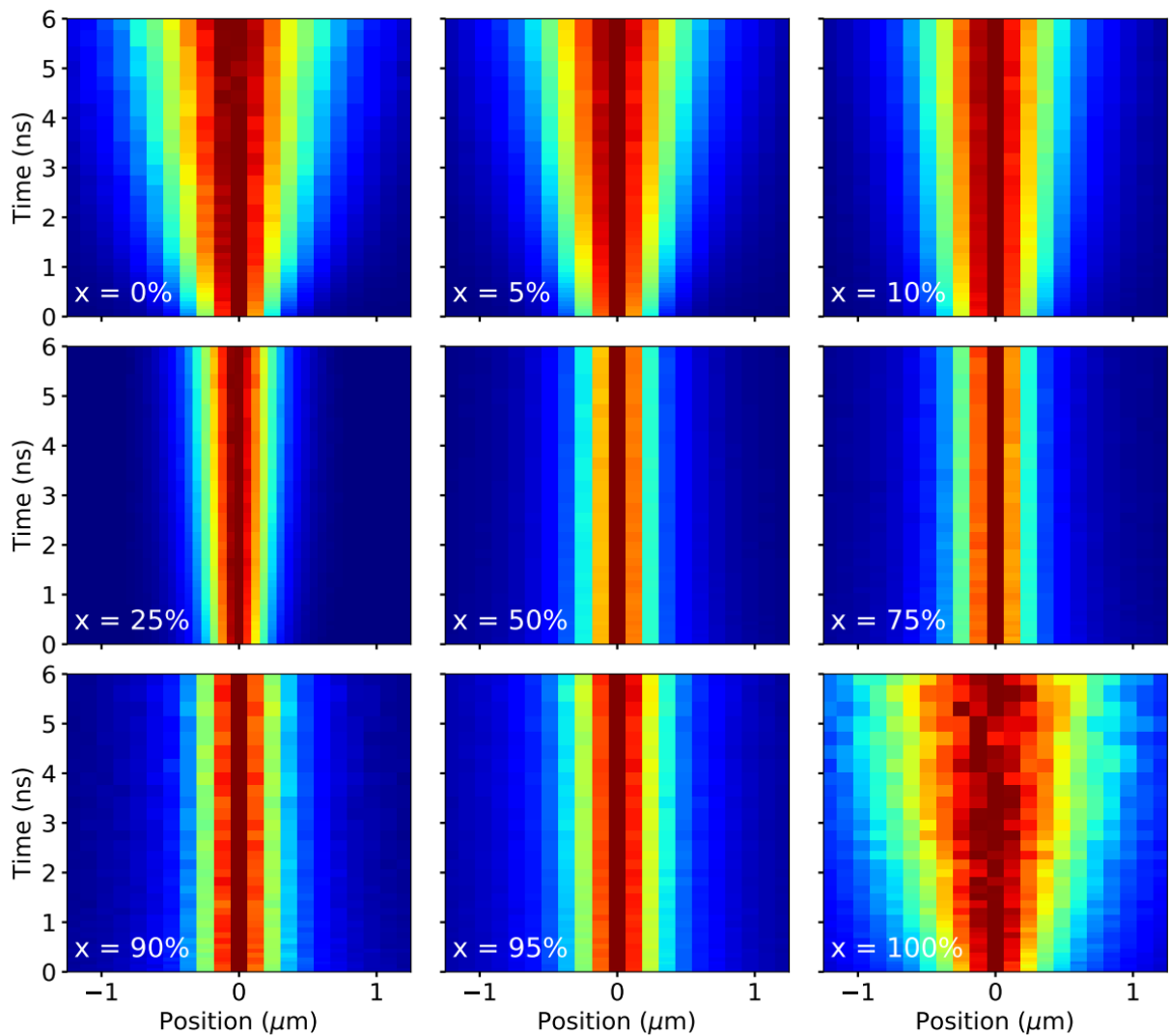

**Supplementary Fig. 3.** Diffusion maps for various mixed-halide perovskites (PEA)$_2$Pb(I$_{1-x}$Br$_x$)$_4$ with x = 0, 5, 10, 25, 50, 75, 90, 95, and 100%.

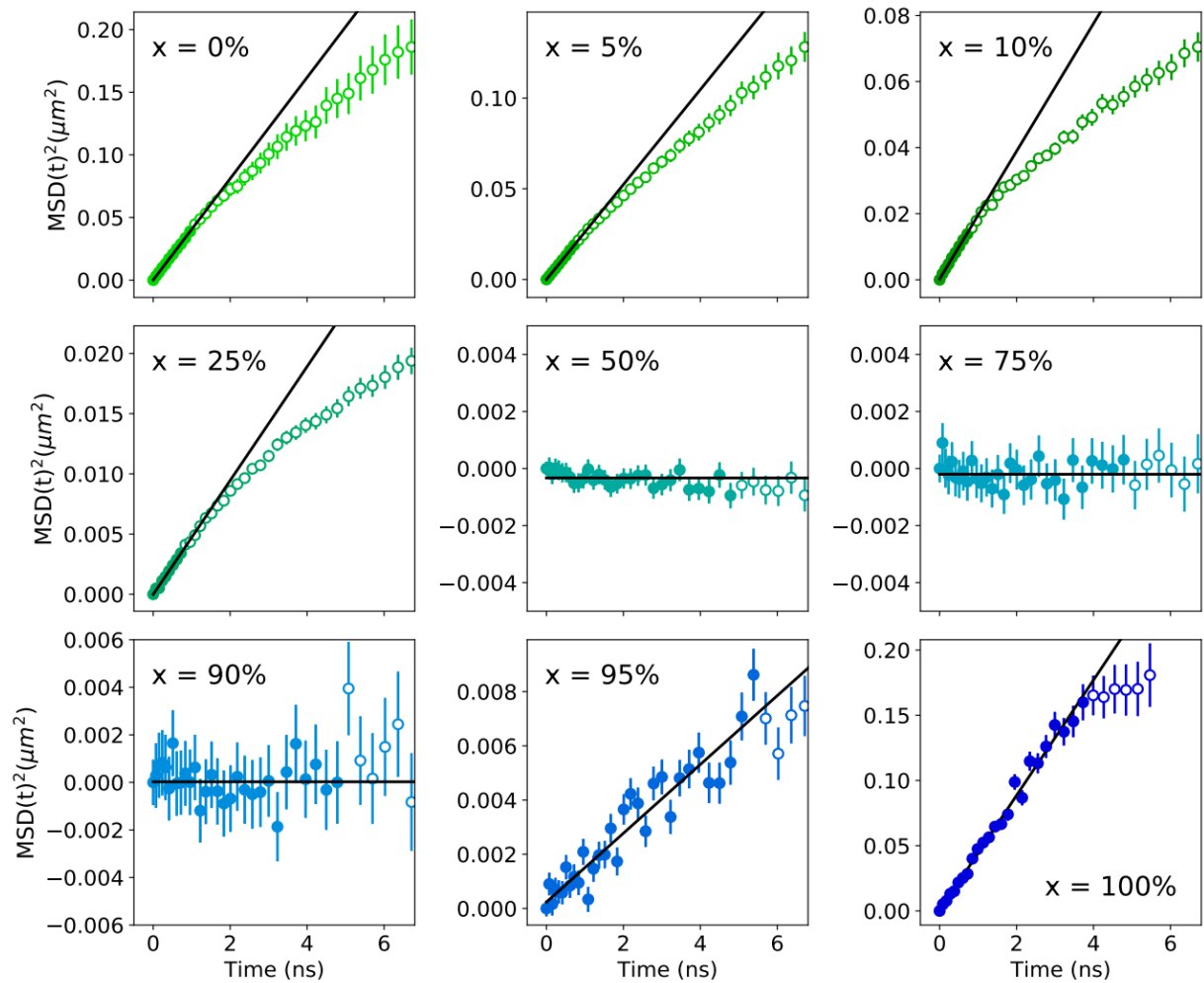

**Supplementary Fig. 4.** Evolution of the mean-square-displacement (MSD) of a near-diffraction limited exciton population in various mixed-halide perovskites (PEA)$_2$Pb(I$_{1-x}$Br$_x$)$_4$ with x = 0, 5, 10, 25, 50, 75, 90, 95, and 100%. Reported errors represent the uncertainty in the fitting procedure for $MSD(t)^2$.

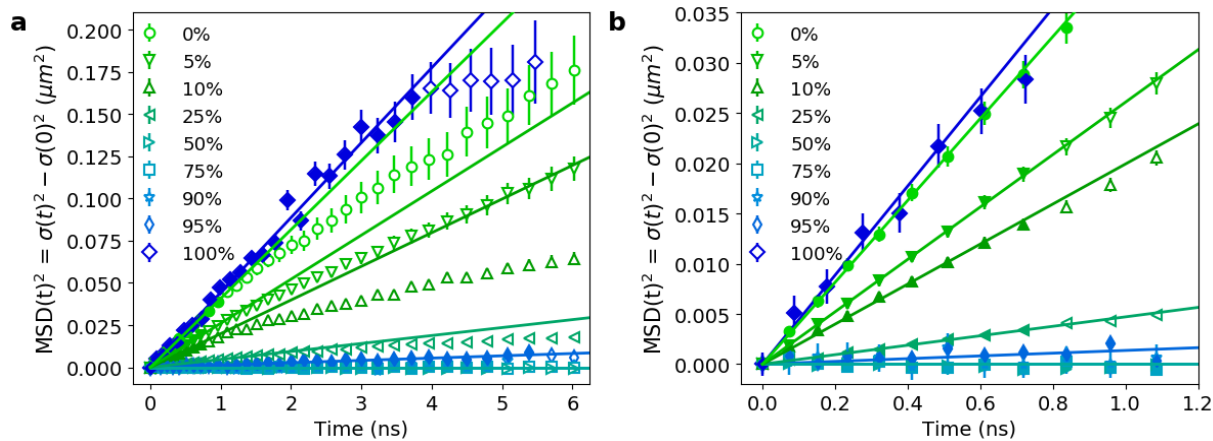

**Supplementary Fig. 5.** Evolution of the mean-square-displacement (MSD) of a near-diffraction limited exciton population in various mixed-halide perovskites (PEA)$_2$Pb(I$_{1-x}$Br$_x$)$_4$ with x = 0, 5, 10, 25, 50, 75, 90, 95, and 100%. **a** Up to 6 ns. **b** Up to 1 ns. Reported errors represent the uncertainty in the fitting procedure for $MSD(t)^2$.

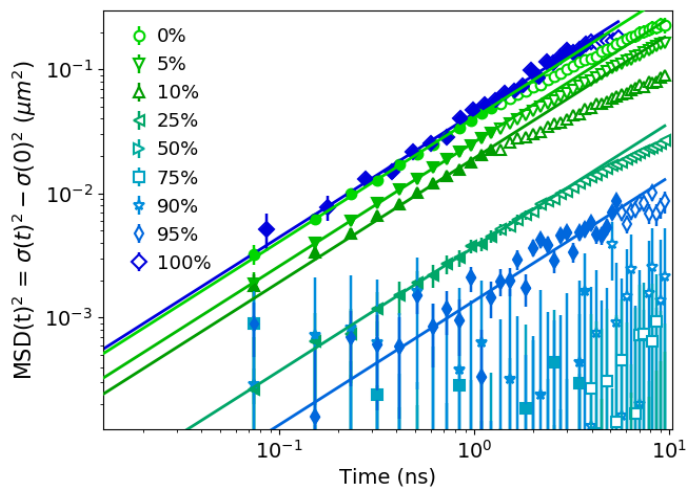

**Supplementary Fig. 6.** Log-log plot of the mean-square-displacement (MSD) of a near-diffraction limited exciton population in various mixed-halide perovskites (PEA)$_2$Pb(I$_{1-x}$Br$_x$)$_4$ with x = 0, 5, 10, 25, 50, 75, 90, 95, and 100%. Reported errors represent the uncertainty in the fitting procedure for $MSD(t)^2$.

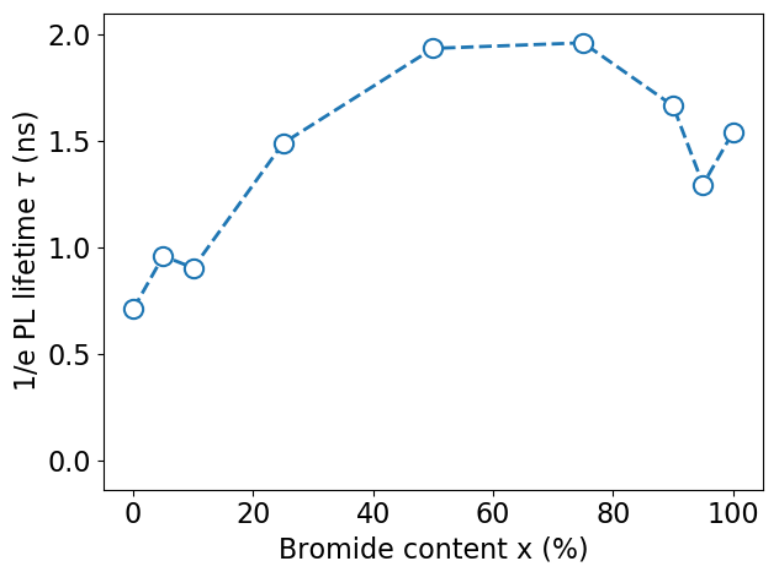

**Supplementary Fig. 7.** Log-log plot of the mean-square-displacement (MSD) of a near-diffraction limited exciton population in various mixed-halide perovskites (PEA)$_2$Pb(I$_{1-x}$Br$_x$)$_4$ with x = 0, 5, 10, 25, 50, 75, 90, 95, and 100%. Reported errors represent the uncertainty in the fitting procedure for $MSD(t)^2$.

For the x = 90, 95, and 100% alloys, part of the emission is blocked by the 420 nm dichroic mirror used in the optical setup. To check the impact of only using part of the emission spectrum for the TPLM measurements, we also performed TPLM with a 50/50 beam splitter instead of a 420 nm dichroic mirror on an x = 100% sample. The challenge with a 50/50 beam splitter is that light of the excitation laser also reaches the detector, which is why we had to attenuate the light reaching the APD, resulting in a reduced S/N of the measurement. To separate the laser light from the sample emission, we used the temporal resolution of TPLM since the laser signal decays faster than the photoluminescence of the perovskite. Looking at the photoluminescence decay measured with the 50/50 beam splitter (orange line in Supplementary Fig. 8) we see that after around 0.6 ns (second dashed line in Supplementary Fig. 8) the emission is dominated by photoluminescence from the perovskite (decay is comparable to photoluminescence measured with the dichroic mirror (blue line)). As a result, we only consider TPLM data after 0.6 ns for the evaluation of the MSD for the measurements with the 50/50 beam splitter (for measurements with the 420 nm dichroic we use data with t > 0 ns). The resulting MSD of the exciton population measured with a 50/50 beam splitter is shown in Supplementary Fig. 9 (please note that t = 0 was shifted by 0.6 ns to have the MSD data start at 0 ns). The observed diffusivity of 0.245 $cm^2s^{-1}$ is comparable with our results obtained from measurements with a 420 nm dichroic mirror (0.222 $cm^2s^{-1}$, Fig. 1 of the main text). As a result, we conclude that the results from TPLM measurements with the 420 nm dichroic mirror represent the diffusivity of the x = 90, 95, 100% samples well, and we use the results from measurements with the dichroic mirror due to the superior S/N ratio of the data.

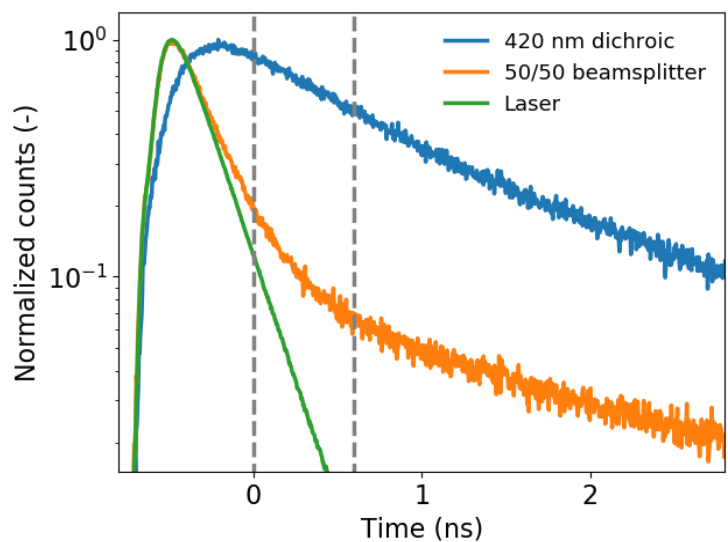

**Supplementary Fig. 8.** Photoluminescence lifetime traces of (PEA)PbBr4 (x = 100%) measured with a 420 nm dichroic mirror (blue line) and a 50/50 beam splitter (orange line). Green line shows the input response function of the excitation laser. With the dichroic mirror (blue line) the influence of the excitation is already negligible at 0 ns (first dashed line). With the 50/50 beam splitter (orange line) the laser excitation light is still significant until around 0.6 ns (second dashed line).

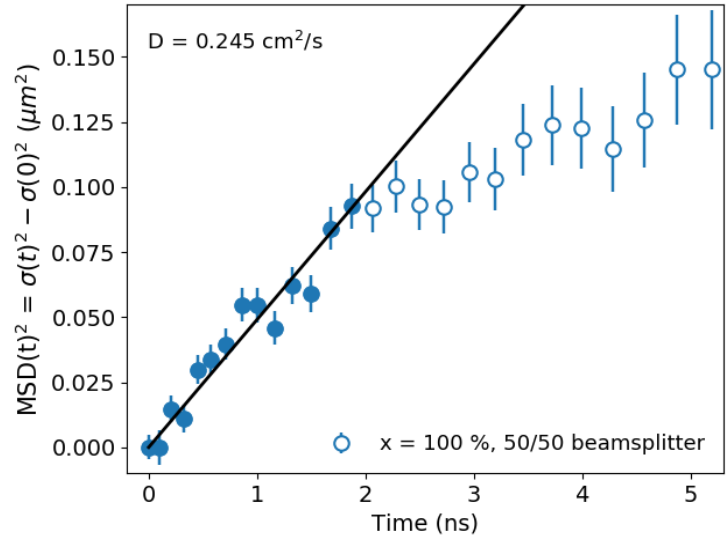

**Supplementary Fig. 9.** Evolution of the mean-square-displacement (MSD) of a near-diffraction limited exciton population in $(PEA)_2PbBr_4$ (x = 100%) measured with a 50/50 beam splitter. Please note that the time axis was shifted by 0.6 ns as compared to Supplementary Fig. 8. Reported errors represent the uncertainty in the fitting procedure for $MSD(t)^2$.

## Supplementary Note 2

**Excitation fluence dependent TPLM.** Perovskites are known to suffer from light induced phase segregation.[2–5] If light induced phase segregation is present in our measurements it should be stronger for higher laser fluences. Therefore, we performed TPLM measurements with two different excitation laser fluences: Our standard 50 nJcm$^{-2}$ and a reduced 5 nJcm$^{-2}$. For this analysis, we focused on the x = 25% sample, which is the sample with the broadest PL emission peak still having a measurable diffusivity. In Supplementary Fig. 10 we show that TPLM measurements with both laser fluences result in the same evolution of the MSD(t) and hence spatial dynamics. The MSD(t) for t > 1ns is slightly lower for 50 nJcm$^{-2}$. However, this is likely caused by the degradation of the perovskite flake (50 nJcm$^{-2}$ was measured after the 5 nJcm$^{-2}$ scan) as traps readily start affecting the later time dynamics.[6] As a result, we exclude photoinduced phase segregation to have a significant impact on the spatial dynamics observed in this study at the laser fluences that were used.

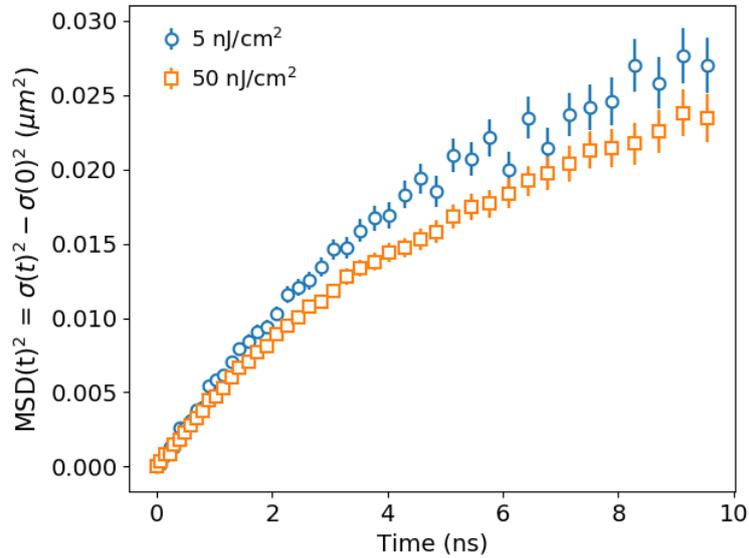

**Supplementary Fig. 10.** Evolution of the mean-square-displacement (MSD) of a near-diffraction limited exciton population in $(PEA)_2Pb(I_{0.75}Br_{0.25})_4$ (x = 25%) measured with two different excitation laser fluences: 5 nJcm$^{-2}$ and 50 nJcm$^{-2}$. Reported errors represent the uncertainty in the fitting procedure for $MSD(t)^2$.

# Supplementary Note 3

**Construction of V(r)**

**Crystal lattice.** We extracted the lattice vectors and coordinates of iodide atoms in a $(PEA)_2PbI_4$ from reported single-crystal XRD data (see Supplementary Table 1).[7] Please note that the out-of-plane coordinate was dropped to project the atoms on a 2D plane, defined by the lattice vectors **a** and **b**. In addition, all other atoms but halides (e.g. Pb) were neglected as they do not have an impact on the local halide composition experienced by the excitons. From $(PEA)_2PbI_4$[7] to $(PEA)_2PbBr_4$[8] the lattice contracts by a factor of (1 – 0.05845). As a result, we approximate the crystal lattices for all mixed-halide perovskites $(PEA)_2Pb(I_{1-x}Br_x)_4$ with the values listed in Supplementary Table 1 for $(PEA)_2PbI_4$ and by scaling the lattice vectors with (1 – 0.05845·x), where x is the bromide content of the mixed-halide perovskite $(PEA)_2Pb(I_{1-x}Br_x)_4$.

**Supplementary Table 1:** Lattice vectors and atom positions of iodide atoms in (PEA)$_2$PbI$_4$ as reported from single crystal XRD measurements.[7]

| Lattice vectors (nm) | **a** = [0.87389, 0] |
| --- | --- |
| | **b** = [0.00544589, 0.8740130] |
| Atom positions [y,z] in terms of the lattice vectors (y***a** + z***b**) | [0.691, 0.8084], [0.4466, 0.4787], [0.0164, 1.02], [0.1905, 0.6906], [-0.0164, 0.98], [0.5534, 0.5213], [0.309, 0.1916], [0.8095, 0.3094] |

**Probability density of 2D excitons.** In 2D systems, the interaction potential between electron and hole is best described by the Keldysh potential.[9,10] Solving the 2D Schrödinger equation yields a solution of the exciton wavefunction $\Psi(R)$, which lies between the 2D hydrogen model ($\Psi(R) \propto e^{-R/a_B}$) and a Gaussian function ($\Psi(R) \propto e^{-R^2/a_B^2}$), where $a_B$ is the exciton Bohr radius and $R$ is the distance from the center of the exciton.[9] For the main text, we decided to use the Gaussian approximation of the exciton with $\Psi(R) \propto e^{-R^2/a_B^2}$ and $|\Psi(R)|^2 \propto e^{-2R^2/a_B^2}$. Supplementary Fig. 11 shows the results of Brownian dynamics simulations for both the Gaussian and the hydrogen model approximation. Simulations with the exact solution to the 2D Schrödinger equation with the Keldysh potential should lie between the two approximations.

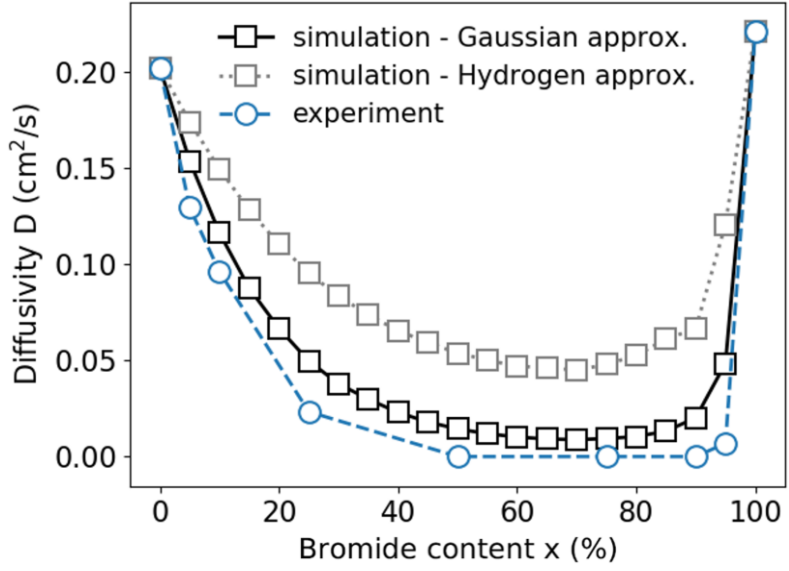

**Supplementary Fig. 11.** Diffusivities obtained from experiments (blue circles) and Brownian dynamics simulations using a Gaussian approximation (black squares) or the hydrogen approximation (gray squares) for the exciton wavefunction to generate the potential landscape $V(x'(r))$. For both approximations, an exciton Bohr radius of $a_B(x) = (1-x) \cdot a_B^I + x \cdot a_B^{Br}$ with $a_B^I = 1.15\ nm$, $a_B^{Br} = 0.7\ nm$, and $r_{cutoff} = 3.5\ nm$ was used.

**Local chemical halide composition x'.** An exciton has a finite size that extends over several unit cells. In 2D, the exciton probability density can be approximated as $|\Psi(R)|^2 \propto e^{-2R^2/a_B^2}$, where $a_B$ is the exciton Bohr radius and $R$ is the distance from the center of the exciton. As a result, we approximate the local chemical composition $x'$ observed by an exciton at position $r$ as a weighted average of bromide and iodide atoms observed by the exciton: $x'(r) = \frac{\sum_{|r_i - r| \leq r_{cutoff}} X_i |\Psi(|r_i - r|)|^2}{\sum_{|r_i - r| \leq r_{cutoff}} |\Psi(|r_i - r|)|^2}$, where $r$ is the exciton position, $r_i$ are the halide atom position, $r_{cutoff}$ is the cutoff radius, $X_i$ is a function that is 1 if the halide atom is bromide and 0 if the atom is iodide. $r_{cutoff}$ was introduced to speed up calculations and was chosen to be more than five times larger than the standard deviation of the probability density $|\Psi(R)|^2$ (here $r_{cutoff} = 3.5\ nm > 5\frac{a_B(x)}{2} \forall a_B(x)$).

**Potential landscape $\Delta V(x'(r))$.** For mixed-halide perovskites (PEA)$_2$Pb(I$_{1-x}$Br$_x$)$_4$, the optical bandgap changes linearly with the Bromide content x.[11] As a result, the bandgap observed by an exciton at position $r$ can be calculated as: $E_g(x'(r)) = (1-x') \cdot E_g^I + x' \cdot E_g^{Br}$, with $E_g^I = 2.400 eV$ and $E_g^{Br} = 3.074 eV$.[11] Brownian dynamics simulations were performed with these energetic landscapes (see Supplementary Fig. 12). For the energy landscape, a constant offset can be introduced without impacting the dynamics. As a result, we use $V(r)$ ($= E_g(x'(r)) - \min[E_g(x'(r))]$), where we subtract the minimal occurring bandgap energy, allowing a better visualization and comparison of the different potential landscapes $V(r)$.

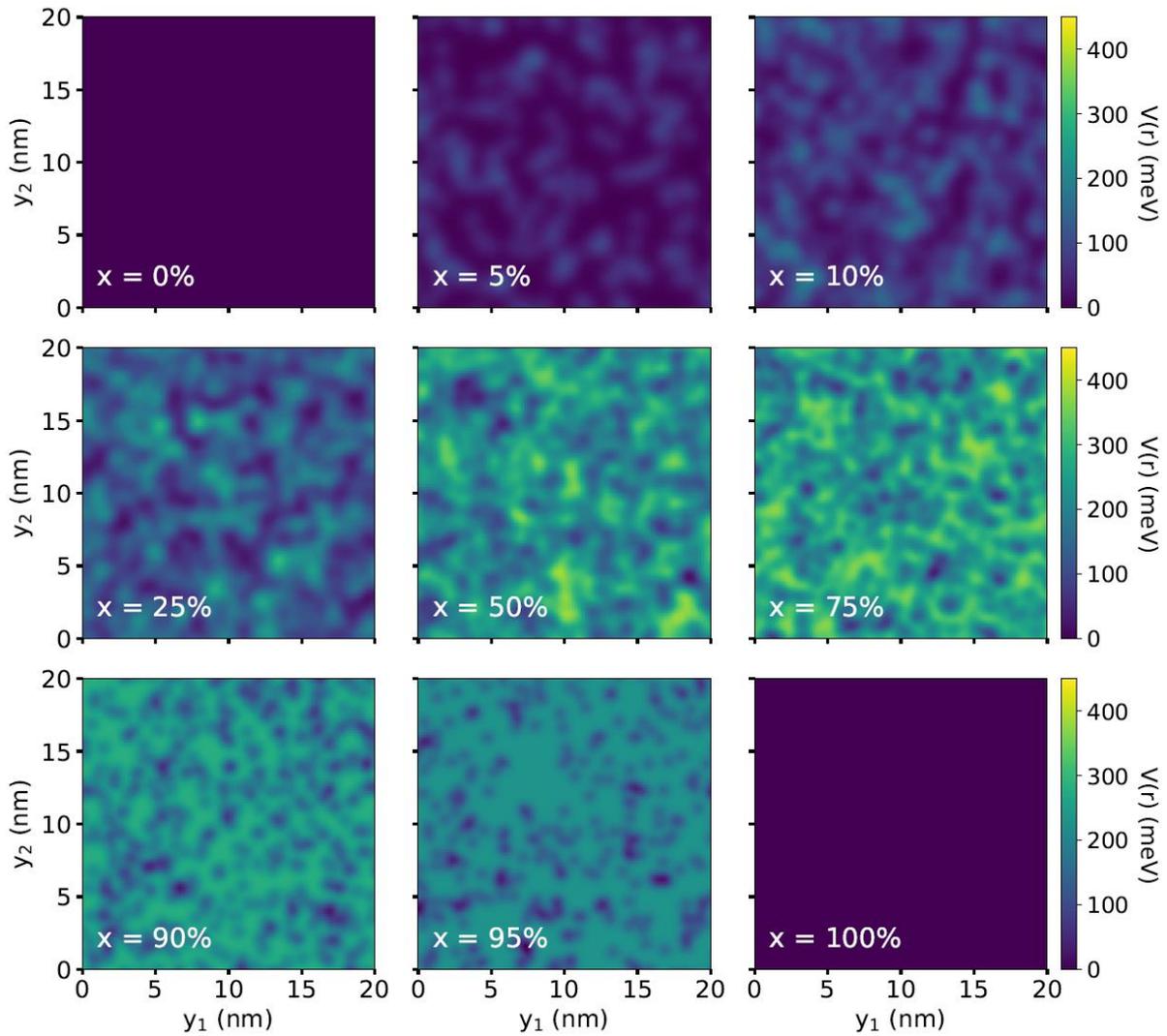

**Supplementary Fig. 12.** Potential landscapes $V(y_1, y_2)$ for different bromide contents x = 0, 5, 10, 25, 50, 75, 90, 95, and 100%.

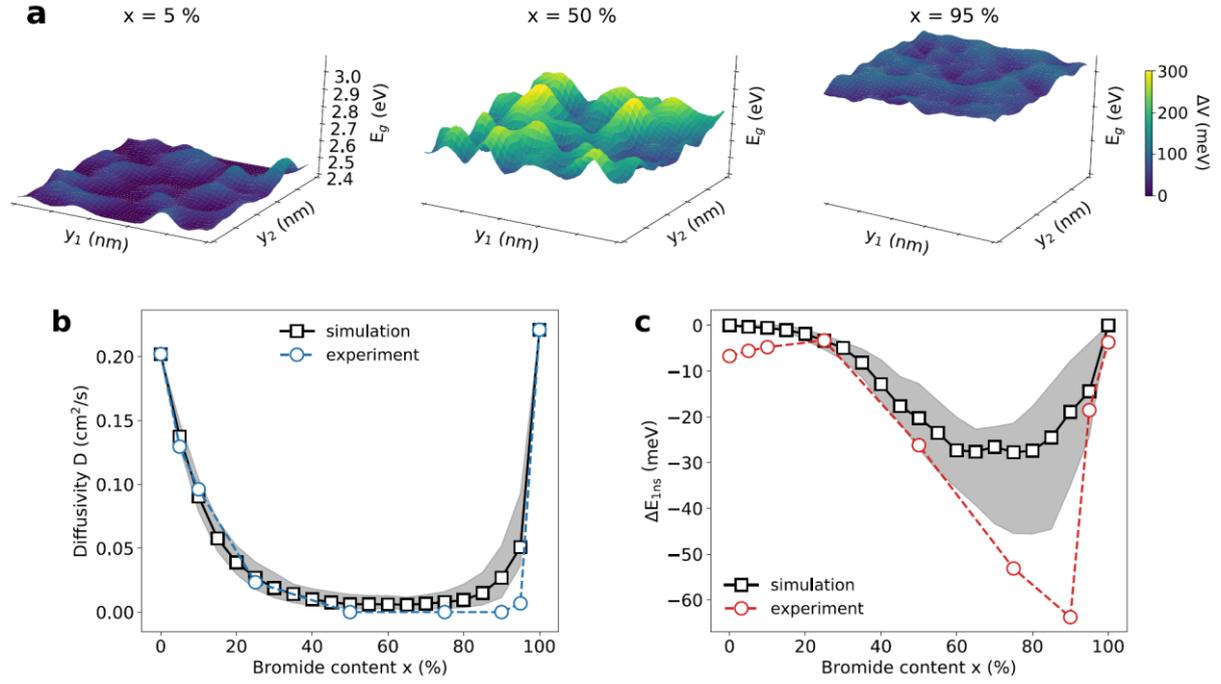

**Supplementary Fig. 13.** Same figure as Fig. 4 of the main text, but simulations were carried out with $a_B^I = a_B^{Br} = 0.8 \pm 0.1\ nm$. **a** Potential landscape for x = 5, 50, 95% for a 10 x 10 nm area. **b,c** Comparison of experiments (open circles) and Brownian dynamics simulations with an exciton Bohr radius of $0.8 \pm 0.1\ nm$ (open squares). It is worth noting that the asymmetry, with a steeper decay on the bromide rich side, is present despite assuming a constant exciton Bohr radius $a_B^I = a_B^{Br} = a_B(x) = 0.8 \pm 0.1\ nm$ for the simulations. We used a moving average $(f(x_i) = \frac{f(x_{i-1})+f(x_i)+f(x_{i+1})}{3})$ for the simulated data points in **c** to reduce the noisiness. **b** Diffusivity $D(x)$ as a function of bromide content x. **c** Change in median energy after 1 ns of photoexcitation as a function of bromide content x.

# Supplementary Note 4

**Brownian dynamics simulations.** We simulated the diffusion process as Brownian walkers in a potential landscape which modeled the interactions between excitons and the crystal structure. The trajectories were represented by the standard stochastic differential motion equations in the Itô interpretation: $\Delta\boldsymbol{r} = \frac{D}{k_B T}\boldsymbol{F}\Delta t + \sqrt{2D_0}d\boldsymbol{W}$, where $\boldsymbol{F} = -\nabla V$ stands for the force felt by an exciton, $D_0$ is the diffusion coefficient, $k_B T$ is the thermal energy and $d\boldsymbol{W}$ is taken from a Wiener process, such that $\langle d\boldsymbol{W}d\boldsymbol{W}\rangle = \Delta t$. Numerical integration was carried out with the straightforward Euler-Maruyama method.

Excitons moved in a two-dimensional 250 x 250 unit cell rectangular simulation box with periodic boundary conditions. As a consequence, the boundaries introduced periodic line crystal defects in the x direction separated by 250 cells in the y direction. We believe these defects had a negligible effect on the simulation results.

At time $t = 0$, excitons were placed radially following a Gaussian distribution of standard deviation $\sigma_0 = 25\ nm$, and the change in the distribution variance was tracked for 2 ns. The program also recorded the evolution of the exciton energy distribution from which we could later calculate the interpolated medians. For each value of the bromide content x, we ran the simulation four times, always generating a new potential landscape for each run and averaging the final results. The one-dimensional MSD was calculated with $MSD(t) = \frac{1}{2}\big(MSD_x(t) + MSD_y(t)\big)$.

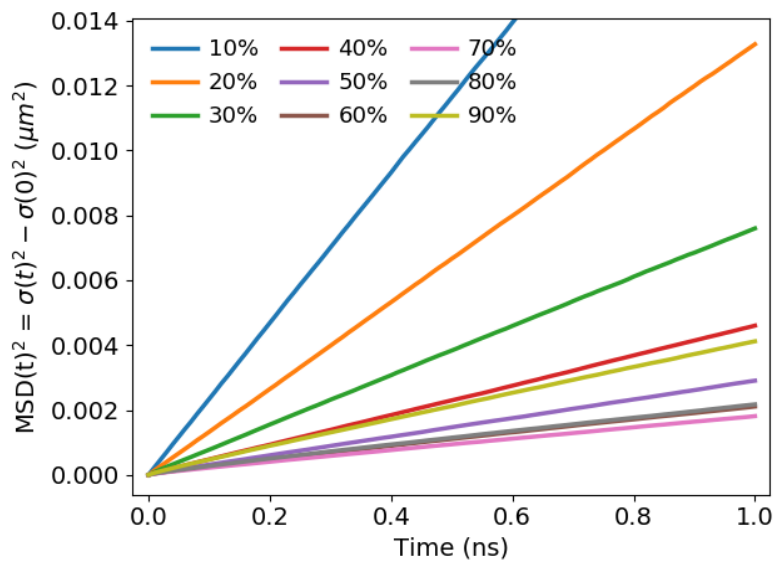

**Supplementary Fig. 14.** Mean-square-displacement (MSD) of the simulations with x = 10, 20, 30, 40, 50, 60, 70, 80, and 90% as shown in Fig. 4b in the main text.

We would like to highlight that the diffusivities $D(x)$ shown in Fig. 4b, Supplementary Fig. 13b, and represented by the slope of the MSD in Supplementary Fig. 14 are effective diffusivities, which are lower than the intrinsic diffusion coefficients $D_0(x)$ that one would observe for a flat energy landscape: $D(x) < D_0(x)$. This is a result of the inhomogeneous energy landscape which restricts the movement of excitons as they get stuck at low-energy sites or move along percolation paths. At very early times, when excitons have not traveled far enough to experience the inhomogeneity of the potential landscape, excitons travel with a diffusivity close to $D_0(x)$.[6,12] However, after less than one picosecond, excitons start experiencing the local disorder resulting in the reduced effective diffusivity $D(x)$ as is shown in Supplementary Fig. 15. The time scale at which this transition happens coincides with the time needed to travel a few nanometers – the characteristic length scale of the inhomogeneous energy landscape ($4\ nm^2/D_0/2 \approx 100\ fs$, see Supplementary Fig. 15). While we are able to resolve these timescales with our simulations, they are, unfortunately, not accessible with our experimental setup. Consequently, we only used $D(x)$ to compare experiments and simulations.

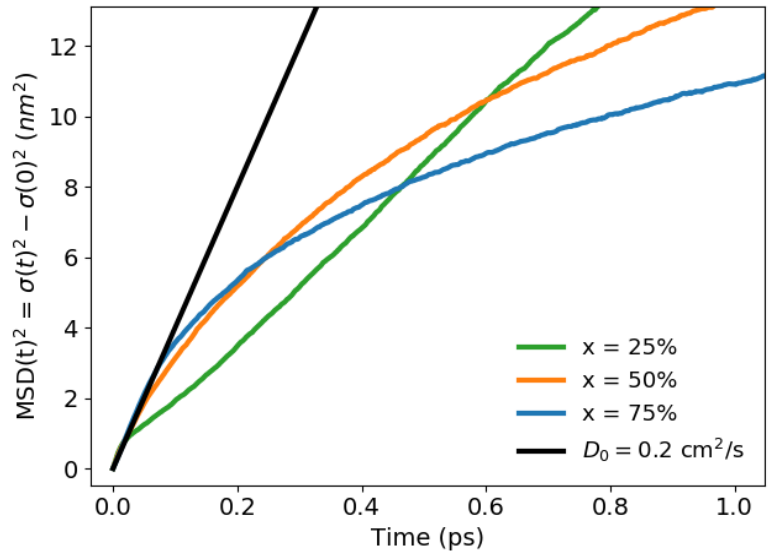

**Supplementary Fig. 15.** Mean-square-displacement (MSD) of the simulations for x = 25, 50, and 75% for early time (≲ 1ps). Showing the transition of a broadening proportional to the diffusion coefficient D0 to a slower effective diffusivity D(x) due to the energetic disorder.

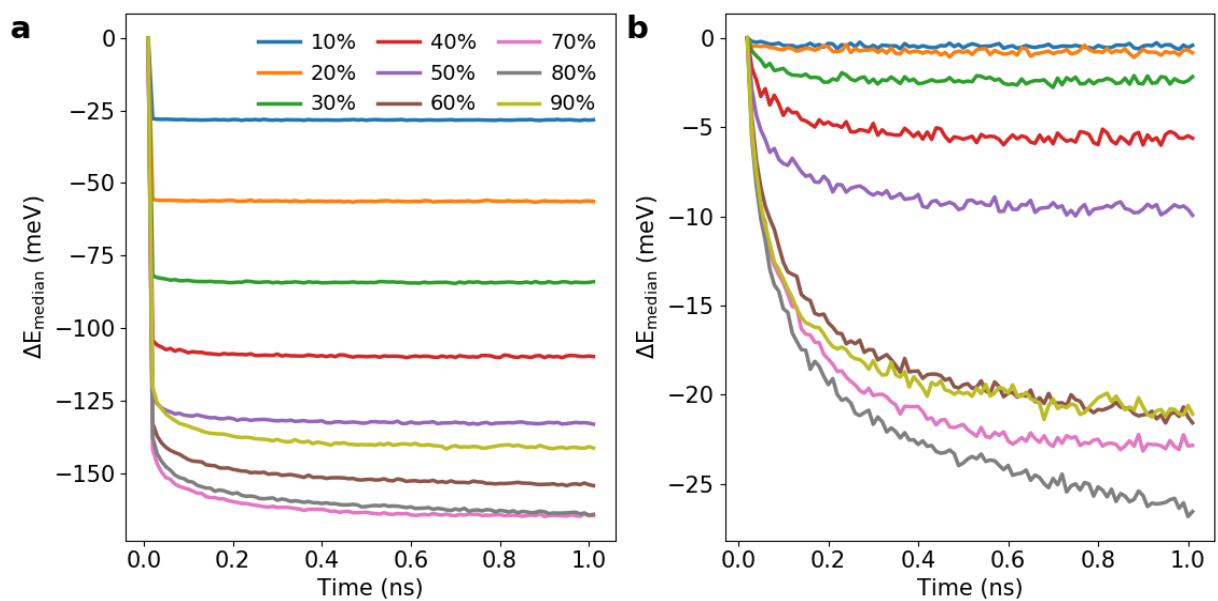

**Supplementary Fig. 16. a** Change of the median energy of the exciton populations shown in Fig. 4c. **b** Same as **a** but without the energy shift within the first 10 ps. Value of $\Delta E_{median}(1\ ns)$ from **b** was taken for the comparison for the experimental data in Fig. 4c.

## Supplementary References

1. Seitz, M. *et al.* Exciton diffusion in two-dimensional metal-halide perovskites. *Nat. Commun.* **11**, (2020).

2. Knight, A. J. *et al.* Electronic Traps and Phase Segregation in Lead Mixed-Halide Perovskite. *ACS Energy Lett.* **4**, 75–84 (2019).

3. Brennan, M. C., Draguta, S., Kamat, P. V. & Kuno, M. Light-Induced Anion Phase Segregation in Mixed Halide Perovskites. *ACS Energy Lett.* **3**, 204–213 (2018).

4. Hoke, E. T. *et al.* Reversible photo-induced trap formation in mixed-halide hybrid perovskites for photovoltaics. *Chem. Sci.* **6**, 613–617 (2015).

5. Bischak, C. G. *et al.* Origin of Reversible Photoinduced Phase Separation in Hybrid Perovskites. *Nano Lett.* **17**, 1028–1033 (2017).

6. Seitz, M. *et al.* Mapping the Trap-State Landscape in 2D Metal-Halide Perovskites Using Transient Photoluminescence Microscopy. *Adv. Opt. Mater.* 2001875 (2021). doi:10.1002/adom.202001875

7. Du, K. Z. *et al.* Two-Dimensional Lead(II) Halide-Based Hybrid Perovskites Templated by Acene Alkylamines: Crystal Structures, Optical Properties, and Piezoelectricity. *Inorg. Chem.* **56**, 9291–9302 (2017).

8. Gong, X. *et al.* Electron-phonon interaction in efficient perovskite blue emitters. *Nat. Mater.* **17**, 550–556 (2018).

9. Prada, E., Alvarez, J. V., Narasimha-Acharya, K. L., Bailen, F. J. & Palacios, J. J. Effective-mass theory for the anisotropic exciton in two-dimensional crystals: Application to phosphorene. *Phys. Rev. B* **91**, 245421 (2015).

10. Keldysh, L. V. Coulomb interaction in thin semiconductor and semimetal films. *J. Exp. Theor. Phys. Lett.* **29**, 658 (1979).

11. Lanty, G. *et al.* Room-Temperature optical tunability and inhomogeneous broadening in 2D-layered organic-inorganic perovskite pseudobinary alloys. *J. Phys. Chem. Lett.* **5**, 3958–3963 (2014).

12. Delor, M. *et al.* Carrier Diffusion Lengths Exceeding 1 µm despite TraLimited Transport in Halide Double Perovskites. *ACS Energy Lett.* 1337–1345 (2020). doi:10.1021/acsenergylett.0c00414